\documentclass[aps, preprint, showpacs, epsfig]{revtex4}
\usepackage{graphics}
\usepackage{graphicx}
\usepackage{epsfig}
\def \be {\begin{equation}}
\def \ee {\end{equation}}
\def \ba {\begin{eqnarray}}
\def \ea {\end{eqnarray}}
\def \bm {\begin{displaymath}}
\def \em {\end{displaymath}}
\def \br {{\bf r}}

\begin{document}
\title{Crystallization of Simple Fluids: Relative Stability of f.c.c. and b.c.c Structures}
\author{Swarn Lata Singh and Yashwant Singh}
\author{}
\affiliation{Department of Physics, Banaras Hindu University, 
Varanasi-221 005,
India}
\date{\today}
\begin{abstract}
A free-energy functional for a crystal that contains both the symmetry conserved and symmetry broken
parts of the direct pair correlation function is developed. The free-energy functional is used to
investigate the crystallization of fluids interacting  via the inverse power potential ;
$u(r)=\epsilon {(\sigma/r)}^n$. In agreement with simulation results we find that for $n=12$ the
freezing is into close packed f.c.c structure while for soft repulsions $(n\leq 6)$ b.c.c phase is more stable.
\end{abstract}
\maketitle
When a fluid freezes into a crystalline solid its continuous symmetry of translation and rotation is broken
into one of the symmetry groups of the Bravais lattices. A crystalline solid has a discrete set of vectors
${\bf R_i}$ such that any function of position, such as one particle density satisfies
$\rho(\br)=\rho(\br+{\bf R_i})$ for all ${\bf R_i}$ [1]. This set of vectors necessarily forms 
a Bravais lattice. While many metallic systems freeze into b.c.c structure, simple fluids like Ar freeze
into f.c.c structure [2]. Model fluids interacting via inverse power potentials $u(r)= \epsilon (\sigma/r)^n$
where $\epsilon$, $\sigma$ and n are potential parameters and r molecular separation, show interesting freezing
transitions [3-5]. The more repulsive ($n \geq 7$) systems freeze into f.c.c structure while the soft
repulsions $n \leq 6$ freeze into b.c.c phase. A correct description of the stability of the two cubic structures
is a stringent test of any statistical mechanical theory based on first principle, since atomic 
arrangements in the two are very different.

Since 1979 [6] the density functional theory (DFT) has been applied to the problem of crystallization of a 
wide variety of fluids [7-8]. Despite its many successes notably with hard sphere system, a full theory applicable
to all intermolecular potentials has remained elusive [7-11]. The direct pair correlation function (DPCF) that
appears in the expression of the free-energy functional corresponds to the ordered phase and is functional
of $\rho (\br)$. When this DPCF is replaced by that of the co-existing isotropic liquid [6] or by that of
an ``effective fluid" [12], the free-energy functional becomes approximate and fails to provide a correct
description of the freezing transition. Attempts to include a term involving three-body direct correlation
function of the isotropic phase in the free-energy functional have failed to improve the situation [9,10].

Due to breaking of symmetry at the fluid-solid transition a qualitatively new contribution [13,14] to the
pair correlation function of a crystal arises . The pair correlation function of a crystal has therefore two
different contributions; one that maintains the continuous symmetry of the Hamiltonian and passes smoothly through 
the transition and the other that breaks it and vanishes at the melting point. In this Letter we develop a method
to estimate the DPCF of a crystal and use it to construct a free energy functional by performing functional
integrations in the appropriate domains of density space. We use this free-energy functional to investigate
the crystallization of fluids interacting via the inverse  power potential. Potentials of this class have a
simple scaling property according to which the reduced excess thermodynamic properties depend on a single
variable which is defined as $\gamma=(\rho \sigma^3)(\epsilon / {k_b}T)^{(3/n)}= {\rho}^*{T^*}^{(-3/n)}$
where $k_b$ is the Boltzmann constant and T, temperature. The equation of state and the melting curve of
these potentials have been extensively investigated by Monte Carlo (MC) simulations for several values
of parameter n so that ``exact" results are available [3-5] for comparison.

The reduced free-energy $A[\rho]$ of an inhomogeneous system is functional of $\rho(\br)$ and is written as 
$A[\rho]=A_{id} [\rho] + A_{ex} [\rho]$ where $A_{id} = \int d\br\rho(\br)[\ln(\rho(\br)\Lambda)-1]$ is
the ideal gas part. Here $\Lambda$ is the cube of thermal wavelength associated with a particle.
The excess part $A_{ex} [\rho]$ is related with the DPCF of the system as [14] 
\ba
\frac {\delta^2A_{ex}[\rho]}{\delta\rho({\bf {r_1}})\delta\rho({\bf {r_2}})}=-c^{(0)}(r,\rho_0)-
c^{(b)}({\bf {r_1}},{\bf {r_2}};[\rho])
\ea
where $c^{(0)}$ is symmetry conserving and $c^{(b)}$ symmetry broken parts of the DPCF. While $c^{(0)}$
depends on the magnitude of the interparticle separation and is function of the average number density
$\rho_0$, $c^{(b)}$ is invariant only under discrete set of translations and is functional of $\rho(\br)$.
If one chooses a center of mass variable 
${\bf r_c}=({{\bf r_1}+{\bf r_2}})/2$ and difference variable $\br={\bf r_1}-{\bf r_2}$,
then $c^{(b)}$ can be written as  [15]; 
\ba
c^{(b)}({\bf r_1},{\bf r_2};[\rho])=\sum_{G}\exp{i({\bf G} \cdot {\bf r_c})}c^{(G)}(\br ;[\rho])
\ea
where ${\bf G}$ are reciprocal lattice vectors (R. L. V.). Since the DPCF is real and symmetric with
respect to interchange of ${\bf r_1}$ and ${\bf r_2}$, 
$c^{( G)}(\br )=c^{(- G)}(\br )$ and  $c^{(G)}(\br )=c^{( G)}(-\br)$

The Ornstein-Zernike equation and the Roger-Young closure relation [16] have been used to calculate
$c^{(0)}(r)$ and $\frac{\partial c^{(0)}(r)}{\partial \rho}$ as a function of intermolecular separation for
densities ranging from zero to a density above the melting point at small intervals for n=12,6 and 4.
This method gives values of pair correlation functions which are thermodynamically self consistent.
For $c^{(b)}({\bf r_1},{\bf r_2},[\rho])$ one can either generalize the integral equation method [14] or
use a functional Taylor expansion [7]. In the latter the contribution due to inhomogeneity to the DPCF
is obtained from the higher-order direct correlation functions of the isotropic phase. The leading term
of this expansion gives
\ba
c^{(b)}({\bf r_1},{\bf r_2};[\rho])=\int c_3^{(0)}(r_{12},r_{13},r_{23};\rho_0)
(\rho({\bf r_3})-\rho_0) d{\bf r_3}
\ea
where $\rho({\bf r_3})=\rho_0+\rho_b({\bf r_3})$ with $\rho_b({\bf r_3})=\rho_0 \sum_{G\not=0}\mu_G
\exp(i {\bf G}\cdot{\bf r_3})$ and $c_3^{(0)}$ is the three-body DPCF of the isotropic fluid of density
$\rho_0$, $\mu_G$ are order parameters and the sum is over the complete set of R.L.V of the crystal.
Eq.(3) satisfies the condition that $c^{(b)}$ is zero in the isotropic
phase and depends on the amplitude and phase factors of density waves which arise in a crystal due
to breaking of symmetry of fluid. Though the higher order terms in (3), particularly for softer spheres,
are not negligible and will affect the location of fluid-solid transition [17], the relative stability 
of the two cubic structures can however be understood on the basis of (3) as the error caused due to
neglect of higher order terms in both structures are of similar magnitude.

Barrat et.al [9] have shown that $c_3^{(0)}$ can be factorized as
$c_3^{(0)}(r_{12},r_{13},r_{23})=t(r_{12})t(r_{13})t(r_{23})$ and the function t(r) can be
determined from the relation 
\ba
\frac{\delta c^{(0)}(r)}{\delta \rho}=t(r)\int t(r_{13})t(r_{23})d{\bf r_3}
\ea
 Using above relations we find the following relation for $c^{(G)}({\bf r})$ :
\ba
c^{(G)}(\br) =\sum_{lm} \lbrack \frac{\rho_0\mu_G}{2\pi^2} 
\sum_{l_1}\sum_{l_2}{i^{(l_1+l_2)}}{(-1)^{l_2}}{\frac{(2{l_1}+1)(2{l_2}+1)}{(2l+1)}}{(C_g(l_1l_2l;000))}^2 \nonumber \\
j_{l_2}(\frac{1}{2}Gr) t(r)B_{l_1}(r,G)Y_{lm}{^\star}(\hat {\bf G})\rbrack Y_{lm}{(\hat \br)} 
\ea

where $C_g$ is the Clebsch-Gordan coefficient,$j_l(x)$ the spherical Bessel function and
\ba
B_{l_1}(r,G)=(4\pi)^2 \int dkk^2t(k)j_{l_1}(kr)\int dr^{\prime} {r^{\prime}}^{2} t(r^{\prime})
j_{l_1}(kr^{\prime})j_{l_1}(Gr^{\prime})\nonumber\\
\ea

The crystal symmetry dictates that l and $l_1+l_2$ are even and for a cubic crystal, $m=0,\pm 4$. If we write 
$c^{({\bf G})}(\br)=\sum_{lm} c^{(G)}_{(lm)}(r)Y_{lm}(\hat \br)$, the expression given in the square
bracket in (5) is the expression for $c_{lm}^{(G)}(r)$. The $c_{lm}^{(G)}(r)$ depends on the order parameter
and on the magnitude of R.L.V.

In Figs. 1 and 2 we compare few harmonic coefficients $c_{lm}^{(G)}(r)$ of the DPCF of f.c.c and b.c.c
structures for n=12 for R.L.V of first and second sets, respectively. We note that the values of
$c_{lm}^{(G)}(r)$ are far from negligible and differ considerably for the two structures; the difference
is both in magnitude and in r dependence.It is this difference that plays crucial role in giving
relative stability to one crystalline structure over the other. As the magnitude of $\bf G$ increases the values
of $c_{lm}^{(G)}(r)$ decreases and after the ninth set of R.L.V values become negligible.

The functional integration of (1) in density space gives $A_{ex}[\rho]$. In this integration the system
is taken from some initial density to the final density $\rho(\br)$ along a path in the density space; 
the result is independent of the path of integration [18]. Since the symmetry conserving part $c^{(0)}$
depends on number density only, the integration is done taking the density of the coexisting fluid $\rho_l$
as reference. This leads to 
\ba
A_{ex}^{(0)}[\rho]=A_{ex}(\rho_l)-{\frac{1}{2}}\int d{\bf r_1}\int d{\bf r_2} 
\Delta \rho({\bf r_1})\Delta \rho({\bf r_2}) \overline c^{(0)}({\bf r_1},{\bf r_2}) 
\ea
where $\overline c^{(0)}({\bf r_1},{\bf r_2})=2\int_0^1d\lambda \int_0^1d\lambda^{\prime}
c^{(0)}(r;\rho_l+\lambda \lambda^{\prime}(\rho_0-\rho_l))$, $\Delta \rho(\br)=\rho(\br)-\rho_l$,
$A_{ex}(\rho_l)$ is the excess reduced free energy of the isotropic fluid of density $\rho_l$ and
$\rho_0=\rho_l(1+\Delta \rho^{\star})$ is the average density of the ordered phase.
Since the functional integrations of $c^{(b)}$ have to be done in the density space specified by 
the order parameter $\mu_G$ and the number density $\rho_0$ we define the path of integration by
two parameters $\lambda$ and $\xi$ which vary from 0 to 1.
The parameter $\lambda$ raises the density from 0 to $\rho_0$ as it varies from 0 to 1 whereas parameter
$\xi$ raises the order parameter from 0 to $\mu_G$ as it varies from 0 to 1. This integration gives 
\ba
A_{ex}^{(b)}=-\frac{1}{2}\int d{\bf r_1} \int d{\bf r_2} \rho_b(\bf r_1)\rho_b(\bf r_2) 
\overline c^{(b)}({\bf r_1},{\bf r_2})
\ea
where
\ba
\overline c^{(b)}({\bf r_1},{\bf r_2})=4\int_0^1 d\xi \xi \int_0^1d\xi^{\prime} 
\int_0^1 d\lambda \lambda \int_0^1 d\lambda^{\prime} c^{(b)}({\bf r_1},{\bf r_2};\lambda \lambda^{\prime}
\rho_0;\xi \xi^{\prime}\mu_G) 
\ea
The free energy functional for a crystal is the sum of $A_{id}$, $A_{ex}^{(0)}$ and $A_{ex}^{(b)}$.

The grand thermodynamic potential defined as $-W=A-\beta \mu \int d\br \rho(\br)$, where $\mu$ 
is the chemical potential, is used to locate transition as it ensures that the pressure and 
chemical potential of two phases remain equal at the transition. The transition point is 
determined by the condition $\Delta W=W_l-W=0$ where $W_l$ is the grand thermodynamic potential of the fluid. 
We calculate the ideal gas part of $\Delta W$ using the Gaussian ansatz [19] for the solid
density $\rho(\br)={(\alpha/\pi)}^{3/2} \sum_{R_i}
\exp(-\alpha{(\br-{{\bf R_i})}^2})$, $\alpha$ being the localization parameter and for the
excess part the Fourier form with $\mu_G=\exp(-G^2/4\alpha)$.
\ba
\frac {\Delta W}{N} = 1-(1+\Delta \rho^{\star})[\frac{5}{2}+\ln \rho_l-\frac{3}{2}\ln{(\alpha/\pi)}]-
\frac{1}{2}\Delta {\rho^{\star}}^2 \hat c^{(0)}(0)-\frac{1}{2}\sum_{{\bf G}\not=0} {|\mu_G|}^2 
\hat c^{(0)}(G) \nonumber \\
- \frac{1}{2} \sum_{\bf G}\sum_{\bf G_1}\mu_{G_{1}}\mu_{-G-{G_1}} \hat {\overline {c}}^{(G)}
({\bf G_1}+\frac{1}{2}{\bf G})
\ea
where $\hat c^{(0)}(G)=\rho_l \int c^{(0)}(r)e^{i{\bf G} \cdot \br}d\br$  and  $\hat {\overline {c}}^{(G)}
({\bf G_1}+{\frac {1}{2}}{\bf G})=\rho_l\sum_{lm}\int{\overline {c}_{lm}}^{(G)}(r)e^{i({\bf G_1}+
\frac{1}{2}{{\bf G}}) \cdot {\br }} Y_{lm}(\hat \br) d\br $

We used the above expression to examine the relative stability of f.c.c and b.c.c structures under 
conditions of fluid-solid coexistence, as determined by MC simulations for soft spheres with n=12, 6 and 4.
In table 1 we give the values of ideal, symmetry conserving and symmetry broken contributions to $\Delta W/N$. 
We note that the contribution arising due to symmetry broken part of DPCF is far from negligible and its
importance increases with the softness of potential. While it is about one-fourth of the symmetry conserving 
part for n=12, for n=4 it increases to nearly one-half. As the contribution is negative it stabilizes
the solid phase. Without it the theory strongly overestimates the stability of fluid phase specially for 
the softer repulsions (n=6 and 4) [11,20]. In agreement with simulation results we find that for
n=12 the f.c.c structure on freezing is favored while for the softer repulsions (n=6 and 4), 
b.c.c structure is favored.
From the values of $\Delta W/N$ given in the table we also conclude that while (3) is a good approximation
for $c^{(b)}$ for $n\geq 12$ but it overestimates the value of  $c^{(b)}$ as the softness
of the repulsion increases. Because of this the crystal phase becomes stable at lower values of $\gamma$
than found by  simulations for $n=6$ and $4$.

In conclusion, we developed a free-energy functional for a crystal that contains both symmetry conserved
and symmetry broken parts of the direct pair correlation function. We calculated the symmetry conserving
part of the pair correlation functions using the Ornestein-Zernike equation and the Roger-Young closure
relation and used a perturbation expansion for $c^{(b)}$. The $c^{(b)}({\bf r_1}, {\bf r_2})$ has been 
expressed in the Fourier series in the center of mass variable with coefficient $c^{(G)}({\br})$ which
is function of difference $\br (={\bf r_1}-{\bf r_2})$ and is found to differ considerably both in
magnitude and in dependence on r for the b.c.c and f.c.c structures. In agreement with simulation
results we find that for n=12 the freezing is into closed packed f.c.c structure while for soft repulsions
($n=6$ and $4$) b.c.c phase becomes more stable. The predictive power of the free-energy functional
developed here can further be improved by improving the accuracy of $c^{(b)}({\bf r_1}, {\bf r_2})$ which
can perhaps be achieved by including one more term in (3) [17]. 

We thank J. Ram and P. Mishra for their help in computation. This work was supported by a research grant
from DST of govt. of India, New Delhi. One of us (SLS) thanks UGC (New Delhi) for research fellowship.

\begin{figure}[h]
\includegraphics[width=3.5in]{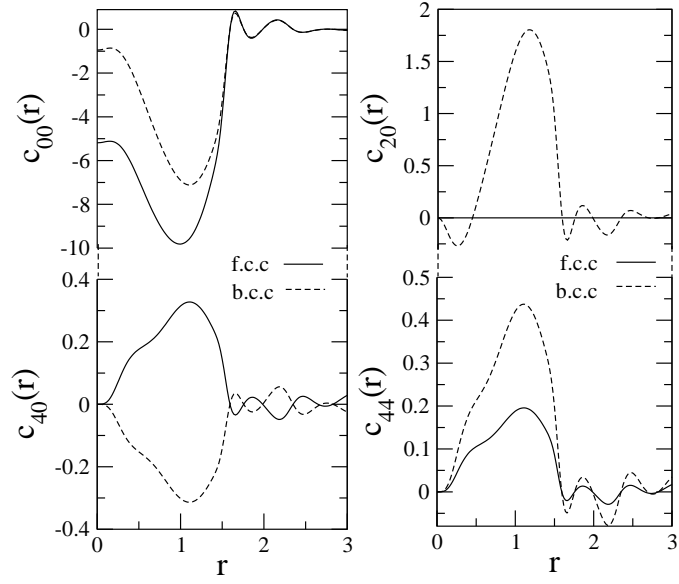}
\caption{Harmonic coefficients ${c^{(G)}}_{lm}(r)$ for a R. L. V. of the first set for $n=12$. } 
\end{figure}

\begin{figure}[h]
\vspace {0.5in}
\includegraphics[width=3.5in]{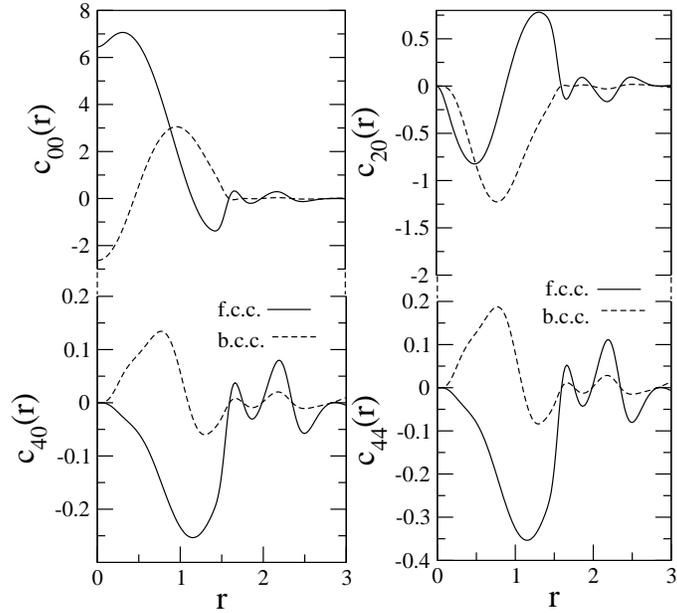}
\caption{Harmonic coefficients ${c^{(G)}}_{lm}(r)$ for a R. L. V. of the second set for $n=12$. } 
\end{figure}
\begin{table*}
\caption{ Ideal $(\Delta W_{id}/N)$, symmetry conserving $(\Delta W_0/N)$ and symmetry broken $(\Delta W_b/N)$ 
contributions to $\Delta W/N$ for f.c.c and b.c.c phases of three inverse power potentials
($n=12, 6$ and $4$) at fluid-solid coexistance. The coexistance parameters $\gamma_s$,
$\gamma_l$, and L (the Lindemann parameter) are also given}
\begin{ruledtabular}
\begin{tabular}{|c|c|c|c|c|c|}
\small
${}$   & structure&$\Delta W_{id}/N$& $\Delta W_0/N$ & $\Delta W_b/N$ &$\Delta W/N$ \\ \hline
$n=12, L=0.15$ &f.c.c &2.80  &-2.29&-0.50 &0.01 \\
$\gamma_s=1.19$, $\gamma_l=1.15$ &b.c.c &2.88  &-2.24 &-0.55 &0.09 \\ \hline
$n=6, L=0.17$ &f.c.c &2.38  &-1.80&-0.62 &-0.04 \\
$\gamma_s=2.33$, $\gamma_l=2.30$ &b.c.c &2.46  &-1.87 &-0.69 &-0.10 \\ \hline
$n=4, L=0.18$ &f.c.c &2.19  &-1.60&-0.72 &-0.13 \\
$\gamma_s=5.75$, $\gamma_l=5.72$ &b.c.c &2.28  &-1.75 &-0.82 &-0.29 \\ 
\end{tabular}
\end{ruledtabular}
\end{table*}

\end{document}